\documentclass[12pt]{article}
\usepackage{graphicx}
\begin{document}

\vbox{\vspace{20mm}}

\begin{center}

{\Large \bf Lorentz Coherence and the Proton Form Factor}\\
\vspace{7mm}
Young S. Kim \\
Center for Fundamental Physics, \\University of Maryland,
 College Park, Maryland 20742, U.S.A.\\
 e-mail: yskim@umd.edu

\end{center}

\vspace{20mm}
\begin{abstract}
The dipole cutoff behavior for the proton form factor has been and still
is one of the major issues in high-energy physics.  It is shown that this
dipole behavior comes from the coherence between the Lorentz contraction
of the proton size and the decreasing wavelength of the incoming photon
signal.  The contraction rates are the same for both cases.  This form of
coherence is studied also in the momentum-energy space.  The coherence
effect in this space can be explained in terms of two overlapping wave
functions.

\end{abstract}

\vspace{60mm}
\noindent to be published in the Special Issue of Physica Scripta
dedicated to 150 Years of Margarita and Vladimir Man'ko.

\newpage
\section{Introduction}
Einstein and Bohr met occasionally to discuss physics.  Einstein
was interested in how things look to moving observers, while Bohr
was interested in the electron orbit of the hydrogen atom.  Thus,
they must have talked about how the orbit looks to a moving observer.
\par
\begin{figure}[thb]
\centerline{\includegraphics[scale=0.4]{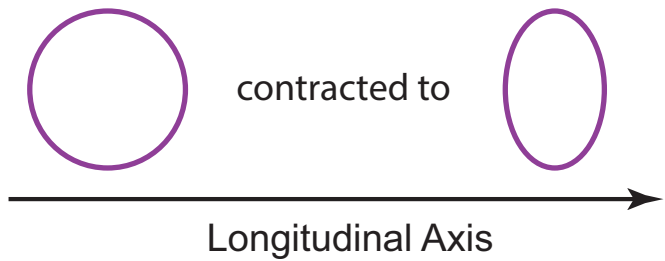}}
\caption{Classical picture of Lorentz contraction.  We expect that
the longitudinal component becomes contracted while the transverse
components are not affected.  The issue is how this picture appears
in the world of quantum mechanics.}\label{locontrac}
\end{figure}

It is possible that they thought about the circular orbit with a
longitudinal contraction as illustrated in Fig.~\ref{locontrac}.  This
picture of Lorentz contraction is still common in the physics
literature~\cite{bell87}.  However, after 1927, the orbit became a
standing wave.  If the standing wave has a rotational symmetry, this figure
could still serve a useful purpose.  In this paper, we shall see how this
figure manifests itself in the Lorentz-covariant quantum world.
\par

\par

In 1953, Hideki Yukawa was interested in constructing harmonic oscillator
wave functions that can be Lorentz-transformed~\cite{yuka53}.  His
primarily interest was in the mass spectrum produced by his
Lorentz-invariant differential equation.  However, at that time, his mass
spectrum did not appear to have anything to do with the physical world.
\par
After witnessing a non-zero charge radius of the proton observed by
Hofstadter and McAllister~\cite{hofsta55}, Markov in 1956 considered
using Yukawa's oscillator formalism for calculating the proton form
factor~\cite{markov56}.  What is the form factor?
\par

However, the constituent particles of the
oscillator wave functions were not defined at that time.  Shortly after
the emergence of the quark model in 1964~\cite{gell64}, Ginzburg and
Man'ko in 1965 considered the relativistic harmonic oscillators for
bound-state quarks~\cite{ginz65}.
\par

Using the same harmonic oscillator wave function given in those earlier
papers, Fujimura, Kiobayashi, and Namiki calculated the form factor,
and concluded it decreases like
\begin{equation}\label{dipole00}
\frac{1}{(\mbox{momentum transfer})^4} ,
\end{equation}
for the proton consisting of three spineless quarks~\cite{fuji70}.
This behavior is called the ``dipole cutoff'' in the physics literature.
\par
This is a very significant result in view of the fact that this behavior
is consistent with what we observe in the real world.  In the same year,
Licht and Pagnamenta derived the same result using the oscillator
Lorentz-contracted wave functions in the Breit coordinate
system~\cite{licht70}.  Their idea was to by-pass the question of the
time-separation variable appearing in the covariant formalism.

\par
In 1971, Feynman, Kislinger, and Ravndal noted that the observed hadronic
mass spectra can be explained in terms of the degeneracies of the
three-dimensional harmonic oscillators~\cite{fkr71}, confirming the earlier
suggestion made by Yukawa in 1953~\cite{yuka53}.  They quoted the paper
by Fujimura {\it et al.}~\cite{fuji70}, but they did not mention Yukawa's
paper.  They did not use the normalizable oscillator wave functions developed
in those earlier papers.

\par
Their basic problem was that they did not handle the time-separation properly.
As the solutions of their Lorentz-invariant differential equation,
they contain the Gaussian factor
\begin{equation}\label{fkrg}
\exp{\left\{-\left(x^2 +y^2 + z^2 - t^2\right)\right\}} .
\end{equation}
This form is Lorentz-invariant but monotonically increases as the
time-separation variable $t$ becomes large.  This does not make any sense
in physics.  Yes, they realize this problem and choose to ignore this
variable.  In so doing they destroy the mathematical consistency of their
paper.
\par
On the other hand Fujimura {\it et al.}~\cite{fuji70} used the Gaussian
form
\begin{equation}\label{fujig}
\exp{\left\{-\left(x^2 + y^2 + z^2 + t^2\right)\right\}} ,
\end{equation}
without recognizing that it was suggested by Yukawa in 1953~\cite{yuka53},
and earlier by Dirac~\cite{dir45}.
This form is normalizable in the $t$ variable, but is not invariant under
Lorentz transformations.  Yet, it can be covariant while this Gaussian
function looks differently to a moving observer.  Then what role does this
$t$ variable play in the covariant formalism?
\par
This problem has been one of my main concerns since I published my first
paper on this subject with Marilyn Noz in 1973~\cite{kn73}.  The solution
to the problem was to formulate the above-mentioned harmonic oscillator
within the framework of Wigner's little groups~\cite{wig39,knp86} which
dictate the internal space-time symmetries of the particles in the
Lorentz-covariant world, while incorporating Dirac's c-number time-energy
uncertainty relation~\cite{dir27} and his instant form of relativistic
dynamics~\cite{dir49}.

\par

According to Wigner's symmetry, the internal space-time symmetry of a
massive particle is like $O(3)$ or three-dimensional rotation group.  Thus,
Feynman {\it et al.} are justified in ignoring the time-separation variable
while concentrating on the three-dimensional space, but they did not do it
properly.  To make the situation worse, they make an apology of using $O(3)$
instead of $O(3,1)$.  They should not have made this apology.
\par
In the Kim-Noz-Oh paper of 1979~\cite{kno79jmp}, we were able to construct
a set of the oscillator wave functions as a representation of Wigner's
little group~\cite{kno79jmp}.  Earlier in 1977, again in collaboration with
Noz, I was able to show that the quark model~\cite{gell64} and the parton
model~\cite{fey69a,fey69b} are two different aspects of one Lorentz-covariant
entity~\cite{kn77par}.  This result was reinforced later by my paper of
1989~\cite{kim89}.

\par
The quark and parton models are applicable to two different limits, namely
the low-speed and high-speed protons.  What happens between those two
limits?  The proton form factor is indeed a case for studying continuous
transition starting from the static proton.
\par
The form factor is a Fourier transformation of the proton density
function within the one-photon exchange picture of electron-proton
scattering.  If the proton wave function is of the Gaussian form as
indicated by the hadronic mass spectra, the form factor should be
a Gaussian function of the momentum transfer.  However, it is not the
case in the real world.  The decrease is slower, and it is a dipole
cutoff as given in Eq.(\ref{dipole00}).
\par
\begin{figure}
\centerline{\includegraphics[scale=0.4]{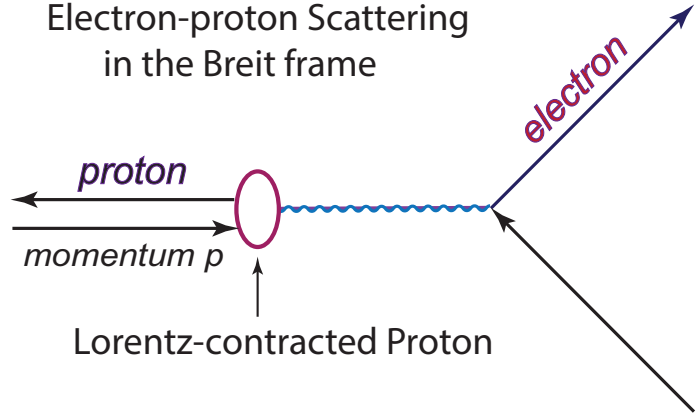}}
\caption{Breit frame.  For the scattering of proton and electron with
the exchange of one photon, it is possible to choose the Lorentz frame
where the momentum of the proton changes to the opposite direction after
the collision.}\label{breit}
\end{figure}

We are thus led to find the resolution of this problem in the relativistic
effect on the size of the proton.  Let us consider the scattering of
electron and proton with one photon exchange in the Lorentz frame as
illustrated in Fig.~\ref{breit}.  This frame is known as the Breit frame
in the literature.

In this collision process, the total momentum is conserved.  The momentum
transfer between the particles becomes the momentum of the photon being
exchanged.   If the momentum transfer becomes larger, the photon wavelength
becomes smaller.
\par
As far as the proton is concerned, it receives an external photon signal.
As its wavelength becomes smaller, the speed of the proton increases
causing a Lorentz contraction of its longitudinal size of the type given
in Fig.~\ref{locontrac}.  We thus suspect
that there is a ``coherence'' between the decrease in the wavelength of the
photon signal and the width of the Lorentz-contracted proton density.
The purpose of this paper is to provide a quantitative analysis for this
Lorentz coherence.

\par
In Sec.~\ref{osc}, we study the Lorentz-contraction property of the
covariant oscillator wave functions.  When boosted, these wave functions
become squeezed along the light cones.  It is shown that this squeeze
leads to Lorentz contraction properties in quantum mechanics.
In Sec.~\ref{ffac}, we calculate the form factor in detail and study the
difference between those with Lorentz-squeezed and with non-squeezed wave
functions. It is noted that, as the momentum transfer increases, the there
is a coherence between the decrease of the wavelength of the incoming photon
signal and the contraction of the width of the proton density.  This
coherence is responsible for the dipole cutoff instead of a steeper
Gaussian cutoff.
In Sec.~\ref{momen}, we study this coherence problem in the momentum-energy
space.  The Lorentz coherence appears as an overlap of two squeezed wave
functions.

\section{Lorentz Contraction of Harmonic Oscillators}\label{osc}

\begin{figure}[thb]
\centerline{\includegraphics[scale=0.5]{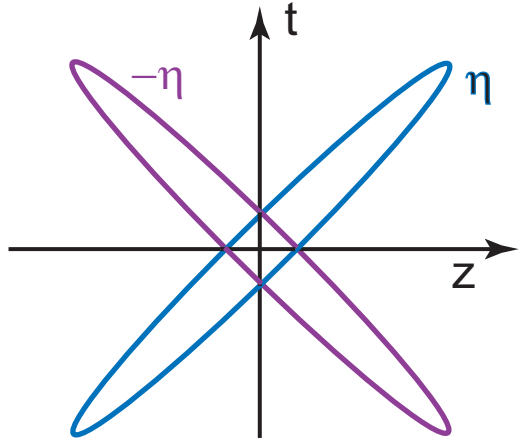}}
\caption{Lorentz-squeezed wave functions. They are squeezed along the
light-cones.  This figure shows two wave functions moving in opposite
directions with the same speed.}\label{overlap11}
\end{figure}

Let us consider two quarks bound together by an oscillator potential.
If we use $x_{\mu}$ for the space-time separation between the quarks,
we can start with the Lorentz-invariant differential equation given
by Feynman {\it et al.}~\cite{fkr71}
\begin{equation}\label{diff33}
\frac{1}{2} \left\{-\left(\frac{\partial}{\partial x_{\mu}}\right)^2
 + x_{\mu}^2 \right\} \psi\left(x_{\mu}\right) = (\lambda + 1)
 \psi \left(x_{\mu}\right) ,
\end{equation}
This equation is separable in the $x, y, z, $ and $t$ coordinates.  If
the hadron moves along the $z$ direction, the transverse components
can be left out.  We can concentrate our attention on the two-variable
equation
\begin{equation}\label{osc05}
\frac{1}{2} \left\{-\left(\frac{\partial} {\partial z}\right)^2
 + \left(\frac{\partial}{\partial t}\right)^2   +
  z^2  -  t^2\right\} \psi (z,t) = \lambda \psi (z,t) .
\end{equation}
This equation has the solution of the form~\cite{yuka53,dir45}
\begin{equation}\label{gauss01}
\frac{1}{\sqrt{\pi}} \exp{\left\{ -\frac{1}{2}\left(z^2 + t^2\right)\right\}} .
\end{equation}
This form is not Lorentz-invariant, but remains localized under the
Lorentz boost
\begin{equation}\label{mat01}
  \pmatrix{z'\cr t'} = \pmatrix{\cosh\eta  &  \sinh\eta \cr
               \sinh\eta  & \cosh\eta}  \pmatrix{z \cr t } ,
\end{equation}
leading to
$$
z \rightarrow z (\cosh\eta) + t (\sinh\eta), \quad\mbox{and}\quad
t \rightarrow z (\sinh\eta) + t (\cosh\eta).
$$
Under this transformation, $\psi(z,t)$ becomes
\begin{equation}\label{gauss02}
\psi_{\eta}(z,t) = \frac{1}{\sqrt{\pi}}
  \exp{\left\{ - \frac{1}{4}\left[e^{-2\eta}(z + t)^2 +
  e^{2\eta} (z - t)^2\right]\right\}} .
\end{equation}
This wave function is squeezed along the light cones as illustrated in
Fig.~\ref{overlap11}.

The present form of quantum mechanics does not allow time-like
excitations~\cite{dir27},  while there are excitations along the $z$
direction.  Thus, the n-th excited-state wave function is~\cite{knp86}
\begin{equation}\label{sol44}
\psi^{n}(z,t) =  \left(\frac{1}{\pi 2^{n}n!} \right)^{1/2}
 H_n(z)\exp{\left\{-\left(\frac{z^2 +t^2}{2}\right)\right\}} ,
\end{equation}
if the hadron is at rest.  If the hadron is Lorentz-boosted,
its Lorentz-boosted wave function should take the form
\begin{eqnarray}\label{sol55}
&{}&\psi_{\eta}^n(z,t) = \left(\frac{1}{\pi 2^{n}n!} \right)^{1/2}
 H_n\left(z\cosh\eta - t\sinh\eta\right)     \nonumber\\[2ex]
&{}&  \hspace{5mm} \times
          \exp{\left\{-\left[\frac{(\cosh2\eta)(z^2 + t^2)
               - 4(\sinh2\eta)zt}{2}\right] \right\}}.
\end{eqnarray}

Let us consider two wave functions with two different values of $n$.
If two covariant wave functions are in the same Lorentz frame
and thus have the same value of $\eta$, the orthogonality relation
\begin{equation}
\left(\psi^{n'}_{\eta}, \psi^{n}_{\eta}\right) = \delta_{nn'}
\end{equation}
is satisfied~\cite{ruiz74}.
If those two wave functions have different values of $\eta$, we
have to start with
\begin{equation}
\left(\psi^{n'}_{\eta'}, \psi^{n}_{\eta}\right) =
\int \left(\psi^{n'}_{\eta'}(z,t)\right)^*\psi^{n}_{\eta}(z,t) dz dt .
\end{equation}
If $\eta'$ = 0, the integral becomes~\cite{ruiz74}
\begin{equation}\label{contrac11}
\left(\psi^{n'}_{0}, \psi^{n}_{\eta}\right) =
\int \left(\psi^{n'}_{0}(z,t)\right)^* \psi^{n}_{\eta}(z,t) dx dt
         =  \left(\frac{1}{\cosh\eta}\right)^{(n + 1)}\delta_{nn'} .
\end{equation}
\par
It is often more convenient to write the $\eta$ parameter in terms
of $\beta = v/c$, where $v$ is the proton velocity.  Then
\begin{equation}\label{contrac33}
        \cosh\eta = \frac{1}{\sqrt{1 - \beta^2}}, \quad
        e^\eta = \sqrt{\frac{1 +\beta}{1 - \beta}} .
\end{equation}
Then the result of Eq.(\ref{contrac11}) can be written as
\begin{equation}\label{contrac22}
\left(\psi^{n'}_{0}, \psi^{n}_{\eta}\right) =
   \left(\sqrt{1 - \beta^2}\right)^{(n + 1)} \delta_{nn'} .
\end{equation}
For the ground state with $n = 0$, it is the Lorentz contraction
factor familiar to us.  For excited state, the explanation is
given in Ref.~\cite{knp86}.

\par

It is then not difficult to write the orthogonality relation
for the non-zero value of $\eta'$, and the result is
\begin{equation}
\left(\psi^{n'}_{\eta'}, \psi^{n}_{\eta}\right) =
   =  \left[\frac{1}{\cosh(\eta -\eta')}\right]^{(n + 1)}\delta_{nn'} .
 \end{equation}
With this understanding, hereafter, we deal only with the ground-state wave
function and drop the superscript.  Thus
\begin{equation}
\left(\psi_{\eta'}, \psi_{\eta}\right) = \frac{1}{\cosh(\eta -\eta')}.
 \end{equation}
Of particular interest is the case with $\eta'= -\eta$, as illustrated in
Fig.~\ref{overlap11}.  This means that
the two oscillator wave functions are moving in opposite directions.
The contraction factor in this case becomes
\begin{equation}\label{contrac55}
  \frac{1}{\cosh(2\eta)} = \frac{1 - \beta^2}{1 + \beta^2} .
\end{equation}
It is interesting to compare this form with that of $1/\cosh(\eta)$ given
in Eq.(\ref{contrac33}).  The difference between these two forms reflect
the  velocity addition law
\begin{equation}
           \frac{2\beta}{1 + \beta^2},
\end{equation}
for two frames moving in opposite directions with the same speed.

\par
In terms of the momentum variable of the moving hadron $p$,
the contraction factors of
Eq.(\ref{contrac33}) and Eq.(\ref{contrac55}) can be written as
\begin{equation}\label{contrac66}
 \frac{1}{\cosh(\eta)} = \frac{1}{\sqrt{1 + p^2}} ,
 \quad\mbox{and}\quad
 \frac{1}{\cosh(2\eta)} = \frac{1}{1 + 2p^2} ,
\end{equation}
respectively.  For simplicity, we use the unit system where the hadronic
mass is one.  We shall use these formulas to study the proton
form factor as a function of $p$ in Sec.~\ref{ffac}.

\section{Proton Form Factors and Lorentz Coherence}\label{ffac}

Without recognizing the papers by Yukawa~\cite{yuka53},
Markov~\cite{markov56}, Ginzburg and Man'ko~\cite{ginz65},
Fujimura {\it et al.}~\cite{fuji70} calculated the electromagnetic form
factor of the proton using the oscillator wave function given in those
earlier papers.  They indeed obtained the desired dipole cutoff.
\par
Also in 1970~\cite{licht70}, Licht and Pagnamenta recognized the problem
with the time-separation variable and carried out the same calculation
for the scattering system in the Breit frame.  In this frame, they were
able to by-pass the dependence on the time-separation variable and were
able to explain the form factor behavior in terms of the Lorentz-contracted
density function.

\par
In the Kim-Noz paper of 1973~\cite{kn73}, we attempted to explain the
form factor in terms of the coherence between the incoming signal and the
width of the contracted wave function.  This aspect
was explained also in terms of the overlap of the energy-momentum wave
function in our book~\cite{knp86}.  In the present paper, I would like to
go back to the coherence problem we raised in 1973~\cite{kn73}, and discuss
the problem in more detail.
\par

We are considering the scattering one electron and one proton by exchanging
one photon.  It is then possible to choose the Lorentz frame in which the
incoming and outgoing protons are moving in opposite directions with the
same speed, as illustrated in Fig.~\ref{breit}.  This Lorentz frame is
known as the Breit frame.
\par
Let us assume that the proton is moving along the $z$ direction as is
indicated in Fig.~\ref{breit}, let $p$ be the magnitude of the momentum,
as in the case of Eq.(\ref{contrac66}).  Then their initial and final
momentum-energy four-vectors are
\begin{equation}
(p, E) \quad\mbox{and}\quad (-p, E) ,
\end{equation}
respectively, where $E = \sqrt{1 + p^2}$.
The momentum transfer in this Breit frame is
\begin{equation}
            (p, E ) - (-p, E) = (2p, 0) ,
\end{equation}
with zero energy component.
\par
The form factor then becomes
\begin{equation}\label{ff01}
F(p) = \int e^{2ipz}
   \left(\psi_{\eta}(z,t) \right)^* \psi_{-\eta}(z,t)~ dz~ dt .
\end{equation}
If we use the ground-state oscillator wave function, this integral becomes
\begin{equation}\label{ff02}
\frac{1}{\pi}
\int e^{2ipz} \exp{\left\{-\cosh(2\eta)\left(z^2 + t^2\right)\right\}
     }~dz~dt .
\end{equation}
The physics of $\cosh(2\eta)$ in this expression was explained in
Eq.(\ref{contrac55}).

\par
In the Fourier integral of Eq.(\ref{ff02}), the exponential function does
not depend on the $t$ variable.  Thus, after the $t$ integration,
Eq.(\ref{ff02}) becomes
\begin{equation}\label{ff05}
F(p) = \frac{1}{\sqrt{\pi \cosh(2\eta)}} \int e^{2ipz}
       \exp{\left\{- z^2\cosh(2\eta)\right\} }~dz .
\end{equation}
If we complete this integral, the form factor becomes
\begin{equation}\label{ff06}
F(p) = \frac{1}{\cosh(2\eta)} \exp{\left\{ \frac{-p^2}
{\cosh(2\eta)} \right\} } .
\end{equation}
If we use the expression of $\cosh(2\eta)$ given in Eq.(\ref{contrac66}),
this form factor becomes
\begin{equation}\label{ff07}
F(p) = \frac{1}{1 + 2p^2} \exp{\left(\frac{-p^2}{1 + 2p^2}\right)} ,
\end{equation}
which decreases as $1/p^{2}$ for large values of $p$.

\par
In order to illustrate the effect of the role of this Lorentz contraction in
more detail, let us perform the integral of Eq.(\ref{ff05}) without the
contraction factor $\cosh(2\eta)$.  This means that the wave function
$\psi_{\eta}(z,t)$ in the Eq.(\ref{ff01}) is replaced by the Gaussian form
$\psi_{0}(z,t)$ of Eq.(\ref{gauss01}).  With this non-squeezed wave function,
the Fourier integral becomes
\begin{equation}\label{gf01}
G(p) = \int e^{2ipz} \left(\psi_{0}(z,t) \right)^* \psi_{0}(z,t)~ dz~ dt.
\end{equation}
The result of this integral is
\begin{equation}\label{gf02}
G(p) = \frac{1}{\sqrt{\pi}} \exp{(-p^2)} .
\end{equation}
This leads to a Gaussian cutoff of the form factor.  This does not happen
in the real world, and the calculation of $G(p)$ is for an illustrative
purpose only.

\par

Let us go back to the Fourier integrals of Eq.(\ref{ff01}) and
Eq.(\ref{gf01}).  The only difference is the $\cosh(2\eta)$ factor
in Eq.(\ref{ff01}).  This factor is in the normalization constant and
comes from the integration over the $t$ variable which does not affect
the Fourier integral.
\par
\begin{figure}
\centerline{\includegraphics[scale=0.5]{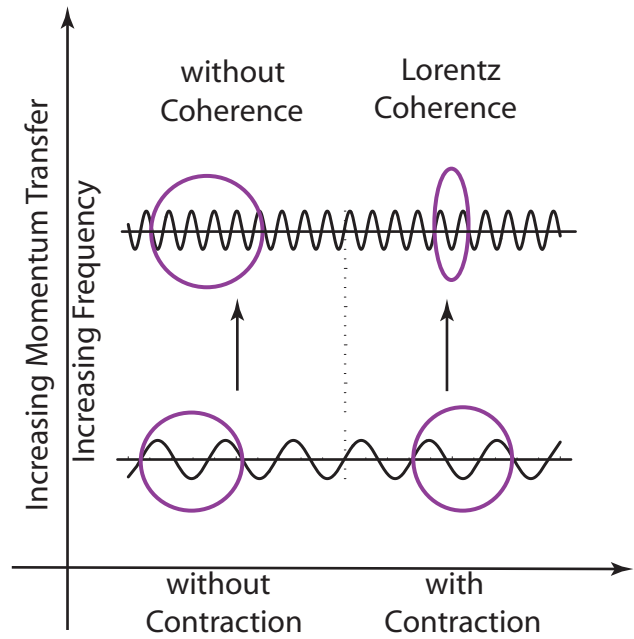}}
\caption{Coherence between the wavelength and the proton size.  Let us go
to Fig.~\ref{breit}, the proton sees the incoming photon.  The wavelength
of this photon becomes smaller for increasing momentum transfer.
If the proton size remains unchanged, there is a rapid oscillation cutoff
in the Fourier integral for the form factor leading to a Gaussian cutoff.
However, it the proton size decreases coherently as the wavelength, the
there are no oscillation effects, leading to a polynomial decrease
of the form factor. }\label{locoher}
\end{figure}

However, it causes the Gaussian width to shrink by $1/\sqrt{2}p$ for large
values of $p$.  The wave length of sinusoidal factor is inversely
proportional to the momentum $2p$.  Thus, both the Gaussian width and
the wavelength of the incoming signal shrink at the same rate of $1/p$
as $p$ becomes large.  Without this coherence, the cutoff is Gaussian
as noted in Eq.(\ref{gf02}).  This effect of this Lorentz coherence is
illustrated in Fig.~\ref{locoher}.

There still is a gap between $F(p)$ of Eq.(\ref{ff07}) and the real world.
Before comparing this expression with the experimental data, we have to
realize that there are three quarks inside the proton with two oscillator
modes as spelled out in the Appendix.
\par
One of the modes goes through the Lorentz coherence process discussed in
this section.  The other mode goes through the contraction process given
in Eq.(\ref{contrac66}).  The net effect is
\begin{equation}
F_{3}(p) = \left(\frac{1}{1 + 2p^2}\right)^{2}
   \exp{\left(\frac{-p^2}{1 + 2p^2}\right)} .
\end{equation}
This will lead to the dsired dipole cutoff of $\left(1/p^2\right)^2$.

\par
In addition, the effect of the quark spin should be addressed.  There
are reports of deviations from the exact dipole cutoff.  There have been
attempts to study the form factors based on the four-dimensional rotation
group with imaginary time coordinate.  There are also many papers based on
the lattice QCD. A brief list of the references to these efforts is given in
my recent paper with Marilyn Noz~\cite{kn11symm}.

\par
The purpose of this paper is limited to studying in detail the role of
Lorentz coherence in keeping the form factor from the steep Gaussian cutoff
in the momentum transfer variable.  The coherence problem is one of the
primary issues of the current trend in physics.

\section{Coherence in Momentum-energy Space}\label{momen}
We are now interested in how the Lorentz coherence manifests itself in
the momentum-energy space.  We can start with the Lorentz-squeezed
wave function in the momentum-energy space, which can be written as
\begin{equation}
\phi_{\eta}\left(q_z, q_0\right) = \frac{1}{2\pi}
   \int e^{-i\left(q_z z - q_0 t \right)}\psi_{\eta}(z,t) dt~dz .
\end{equation}
This is a Fourier transformation of the Lorentz-squeezed wave function of
Eq.(\ref{gauss02}), where $q_z$ and $q_0$ are Fourier conjugate variables
to $z$ and $t$ respectively.  The result of this integral is
\begin{equation}\label{gauss03}
\phi_{\eta}\left(q_z, q_0\right) =\frac{1}{\sqrt{\pi}}
\exp{\left\{-\frac{1}{4}\left[e^{-2\eta}\left(q_z + q_0\right)^2
  + e^{2\eta}\left(q_z - q_0 \right)^2\right]\right\}} .
\end{equation}

\par
In terms of this momentum-energy wave function, the form factor
of Eq.(\ref{ff01}) can be written as
\begin{equation}\label{ff10}
F(p) = \int \phi_{-\eta}^{*}\left(q_0, q_z - p\right)
 \phi_{\eta}\left(q_0, q_z + p\right) dq_0~dq_z .
\end{equation}
The evaluation of this integral leads to the form factor $F(p)$ given
in Eq.(\ref{ff07}).

\par
\begin{figure}[thb]
\centerline{\includegraphics[scale=0.4]{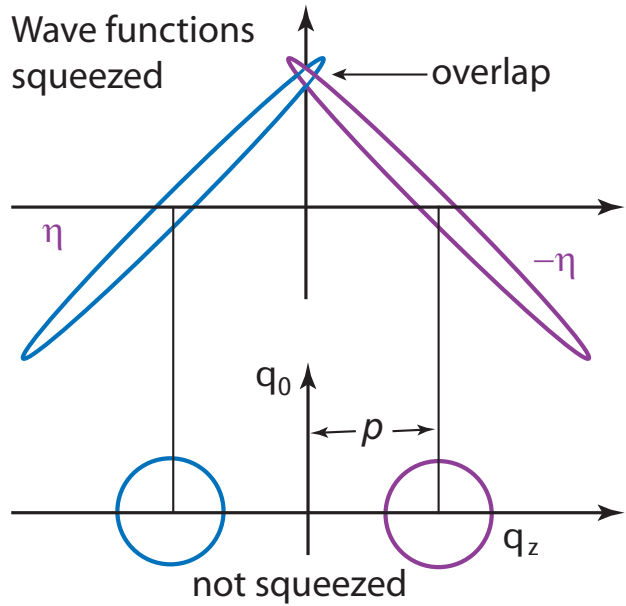}}
\caption{Lorentz coherence in the momentum-energy space.  Both squeezed
and non-squeezed wave functions are given.  As $p$ increases, the two wave
functions in Eq.(\ref{ff10}) become separated.  Without the squeeze, there
are no overlaps.  This leads to a Gaussian cutoff.  The squeezed wave
functions maintain an overlap, leading to a slower polynomial
cutoff.}\label{overlap22}
\end{figure}

In order to see the effect of the Lorentz coherence, let us look at
two wave functions in Fig.~\ref{overlap22}.  The integral is carried
over the $q_{z}~q_{o}$ plane.  As the momentum $p$ increases, the two
wave functions become separated.  Without the Lorentz squeeze, the wave
functions do not overlap, and this leads to a sharp Gaussian cutoff as
in the case of $G(p)$ of Eq.(\ref{gf02}).
\par
On the other hand, the squeezed wave functions have an overlap as show
in Fig.~\ref{overlap22}, and this overlap becomes smaller as $p$ increases.
This leads to a slower polynomial cutoff~\cite{knp86,kn11symm}.

\section*{Conclusions}
Hofstadter's discovery of the non-zero size of the proton opened a new era
of physics~\cite{hofsta55}.  The proton is no longer a point particle.
One way to measure its internal structure is to study the proton-electron
scattering amplitude with one photon exchange, and its dependence on the
momentum transfer.  The deviation from the case with the point-particle
proton is called the proton form factor.

\par
In the experimental front, the dipole cutoff has been firmly established.
Yes, there are also experimental results indicating deviations from this
dipole behavior~\cite{rob05,mat05}.  However, in the present paper, no
attempts have been made to review all the papers written on the corrections.
From the theoretical point of view, those deviations are corrections
from the basic dipole behavior.

While the study of the form factor is still a major subject in physics, it
is gratifying to note that its dipole cutoff comes from the coherence between
the Lorentz contraction of the proton's longitudinal size and the decrease
in the wavelength of the incoming signal.

\section*{Acknowledgments}

Inspired by the paper by Yukawa, Fujimura {\it et al.}, and Feynman
{\it et al.}, I published my first paper on the Lorentz-covariant harmonic
oscillators in 1973 with Marilyn Noz~\cite{kn73}.  Shortly after this
paper appeared in the Physical Review D, Moisey Markov sent me a reprint
of his 1956 review paper published in the Nuovo Cimento~\cite{markov56}.
\par

\begin{figure}[thb]
\centerline{\includegraphics[scale=1.8]{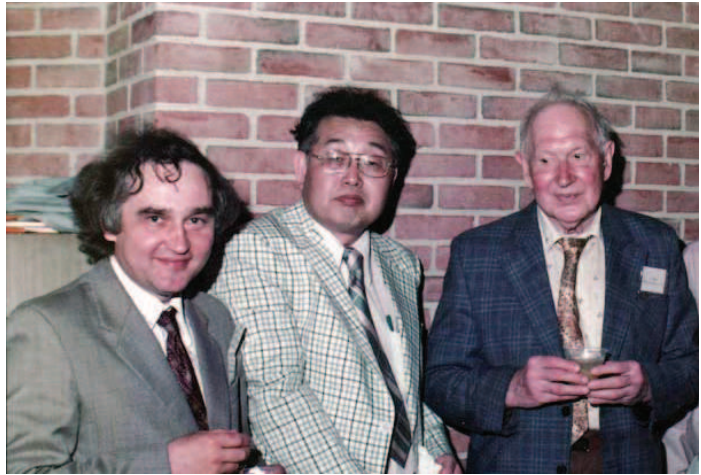}}
\caption{V. I. Man'ko, Y. S. Kim, and M. A. Markov in 1984 during the 13th
International Colloquium on Group Theoretical Methods in Physics held
at the University of Maryland.}\label{markov}
\end{figure}

In 1980, I met Vladimir Man'ko in Cocoyoc (Mexico) while attending the 9th
International Colloquium on Group Theoretical Methods in Physics.  He told
me he saw my papers on the covariant oscillators and told me about his paper
with Ginzburg~\cite{ginz65}.  In 1984, Vladimir Man'ko came with Moisey
Markov to the University of Maryland to attend the 13th the meeting of the
same conference.   During this conference I had a photo with them as shown
in Fig.~\ref{markov}.
\par
In 1986, a graduate student named Yan-Hua Shih of the University of Maryland
told me about Horace Yuen's paper on two-photon coherent states~\cite{yuen76}.
He told me the mathematical formulas in Yuen's paper are very similar to
those in my papers on the covariant harmonic oscillators.  After examining
the paper, I became convinced that the underlying mathematical language
for squeezed states of light is that of the Lorentz group.  In order to
learn more about the subject, I organized in 1991 a workshop on squeezed
states at the University of Maryland.

\par
To this conference, I invited five Soviet physicists, including Vladimir
Man'ko, Margarita Man'ko, and Victor Dodonov.  They became so happy with
this conference that they decided to have the second meeting in 1992 at the
Lebedev Institute in Moscow.  This is how the first meeting on squeezed
states became developed into the international conference series known
to young physicists these days as the ICSSUR.

\par
I am indeed grateful to Victor Dodonov for inviting me to contribute this
paper to this volume dedicated Vladimir Man'ko and Margarita Man'ko on
their 75th birthdays.  I am fortunate enough to use this occasion to
mention Vladimir's early contribution to the subject of harmonic oscillators
in the Lorentz-covariant world.
\par
Finally, I am grateful to both Margarita and Vladimir for their ever-lasting
cooperation and friendship.

\par
\section*{Appendix}

Throughout this paper, we used the hadronic system consisting of two quarks
bound together by a harmonic oscillator force.  The proton however consists
of three quarks.  In this appendix, we explain how this three-body system
becomes that for two oscillators.
\par
This problem was worked out in detail in the 1971 paper of Feynman
{\it et al.}~\cite{fkr71}.  We choose here to use their notation for the
the three quarks.  They use $u_{a}, u_{a}, u_{c}$ for the space-time
coordinates for those quarks.  If there is the oscillator force between the
two quarks, we are led to consider the Gaussian form
\begin{equation} \label{appen01}
  \exp{\left\{ -\frac{a}{2}\left[ \left(u_{a} - u_{b}\right)^{2} +
  \left(u_{b} - u_{c}\right)^{2} + \left(u_{c} - u_{a}\right)^{2}\right] \right\} }.
\end{equation}
In order to deal with this form, Feynman {\it et al.} introduced the
following three variables.
\begin{eqnarray}\label{appen02}
&{}& R = \frac{u_{a} +  u_{b} + u_{a}}{3}, \nonumber \\[1ex]
&{}& x = \frac{u_{b} +  u_{c} - 2 u_{a}}{4},  \nonumber \\[1ex]
&{}& y = \frac{u_{c} - u_{b}}{2\sqrt{3}} ,
\end{eqnarray}
and
\begin{eqnarray}\label{appen03}
&{}& u_{a} = R - 2x, \nonumber \\[1ex]
&{}& u_{b} = R + x - \sqrt{3} y,  \nonumber \\[1ex]
&{}& u_{c} = R + x + \sqrt{3} y .
\end{eqnarray}
In terms of the variables $x$ and $y$, the Gaussian function of
Eq.(\ref{appen01}) becomes
\begin{equation}
\exp{\left\{-\frac{18a}{2}\left(x^{2} + y^{2}\right)\right\}}  .
\end{equation}
This form does not depend on the coordinate variable $R$.  Thus,
the Gaussian form Eq.(\ref{appen01}) for the three-body system becomes
that for two oscillators.

\par


\begin{thebibliography}{99}

\bibitem{bell87}
 Bell J S 1987 {\it Speakable and Unspeakable in Quantum Mechanics:
 Collected Papers on Quantum Philosophy} (Cambridge: Cambridge University)

\bibitem{yuka53}
 Yukawa H 1953 Structure and Mass Spectrum of Elementary Particles I.
 General Considerations {\it Phys. Rev.} {\bf 91} 415-416

\bibitem{hofsta55}
 Hofstadter R and McAllister R W Electron Scattering from the
 Proton 1955  {\it Phys. Rev.} {\bf 98} 217-218

\bibitem{markov56}
 Markov M On Dynamically Deformable Form Factors in the Theory
 of Particles 1965 {\it Suppl. Nuovo Cimento} {\bf 3} 760-772

\bibitem{gell64}
 Gell-Mann M 1964 Nonleptonic Weak Decays and the Eightfold Way
 {\it Phys. Lett.} {\bf 12} 155-156

\bibitem{ginz65}
 Ginzburg V L and Man'ko V I 1965 Relativistic Oscillator Models
 of Elementary Particles {\it Nucl. Phys.} {\bf 74} 577-588

\bibitem{fuji70}
 Fujimura K  Kobayashi, T and Namiki M 1970 Nucleon
 Electromagnetic Form Factors at High Momentum Transfers in an
 Extended Particle Model Based on the Quark Model
 {\it Prog. Theor. Phys.} {\bf 43} 73-79

\bibitem{licht70}
 Licht, A L and Pagnamenta A 1970 Wave Functions and Form Factors
 for Relativistic Composite Particles I. {\it Phys. Rev.} D
 {\bf 2} 1150-1156

\bibitem{fkr71}
 Feynman R P, Kislinger M and Ravndal F (1971)
 Current Matrix Elements from a Relativistic Quark Model
 {\it Phys. Rev.}  D {\bf 3} 2706-2732

\bibitem{dir45}
 Dirac P A M 1945  Unitary Representations of the Lorentz Group
 {\it Proc. Roy. Soc. (London)} A {\bf 183} 284-295

\bibitem{kn73}
 Kim Y S and Noz M E 1973 Covariant harmonic oscillators
 and the quark model {\it Phys. Rev.} D {\bf 8} 3521-3627

\bibitem{wig39}
 Wigner, E. 1939 On Unitary Representations of the Inhomogeneous
 Lorentz Group {\it Ann. Math.} {\bf 40} 149-204

\bibitem{knp86}
 Kim Y S Noz M E 1986 {\it Theory and Applications of the Poincar\'e
 Group} (Dordrecht, The Netherlands: Reidel)

\bibitem{dir27}
 Dirac P A M 1927 The Quantum Theory of the Emission and Absorption of
 Radiation {\it Proc. Roy. Soc. (London)}
 {\bf A114} 243-265

\bibitem{dir49}
 Dirac P A M 1949 Forms of Relativistic Dynamics {\it Rev. Mod. Phys.}
 {\bf 21} 392-399

\bibitem{kno79jmp}
 Kim Y S, Noz M E and Oh S H 1979 Representations of the
 Poincar\'e group for relativistic extended hadrons
 {\it J. Math. Phys.} {\bf 20} 1341-1344

\bibitem{fey69a}
 Feynman R P 1969 Very High-Energy Collisions of Hadrons
 {\it Phys. Rev. Lett.} {\bf 23} 1415-1417

\bibitem{fey69b}
 Feynman R P 1969 {\it The Behavior of Hadron Collisions at Extreme Energies
 in High-Energy Collisions, Proceedings of the Third International
 Conference, Stony Brook, NY} Yang, C. N.{\it et al.} Eds.
 (New York: Gordon and Breach) 237-249.

\bibitem{kn77par}
 Kim Y S and Noz M E (1977) Covariant harmonic oscillators and the
 parton picture {\it Phys. Rev.} D {\bf 15} 335-338


\bibitem{kim89}
 Kim Y S 1989 Observable gauge transformations in the parton
 picture {\it Phys. Rev. Lett.}  {\bf 63} 348-351


\bibitem {ruiz74}
 Ruiz M J 1974 Orthogonality relations for covariant harmonic
 oscillator wave functions {\it Phys. Rev.}  D {\bf 10} 4306-4307


\bibitem{kn11symm}
 Kim Y S and Noz M E 2011 Lorentz Harmonics, Squeeze Harmonics and
 Their Physical Applications {\it Symmetry} {\bf 3}  16-36

\bibitem{yuen76}
 Yuen H P 1976 Two-photon coherent states of the radiation field
 {\it Phys. Rev.} A  {\bf 13} 2226-2243

\bibitem{rob05}
 Alkofer R,  Holl A, Kloker M, Karssnigg A, and Roberts C D 2005
 On Nucleon Electromagnetic Form Factors
 {\it Few-Body Sys.}  {\bf 37} 1-31

\bibitem{mat05}
 Matevosyan H H, Thomas, A W, Miller, G A 2005
 Study of lattice QCD form factors using the extended
 Gari-Krumpelmann model {\it Phys. Rev. C} {\em 72} 065204-5

\end{thebibliography}
\end{document}